\documentclass[twocolumn]{aastex63}

\def\feh{$\mathrm{[Fe/H]}$}

\usepackage{amsmath}

\hypersetup{linkcolor=red,citecolor=blue,filecolor=green,urlcolor=magenta}

\begin{document}

\shorttitle{RR Lyrae in Virgo III}
\shortauthors{Ngeow \& Bhardwaj}

\title{Discovery of RR Lyrae in the Ultra-Faint Dwarf Galaxy Virgo III}

\correspondingauthor{Chow-Choong Ngeow}
\email{cngeow@astro.ncu.edu.tw}

\author[0000-0001-8771-7554]{Chow-Choong Ngeow}
\affil{Graduate Institute of Astronomy, National Central University, 300 Jhongda Road, 32001 Jhongli, Taiwan}

\author[0000-0001-6147-3360]{Anupam Bhardwaj}
\affil{Inter-University Center for Astronomy and Astrophysics (IUCAA), Post Bag 4, Ganeshkhind, Pune 411 007, India}

\begin{abstract}

  Virgo III is a newly discovered ultra-faint dwarf (UFD) candidate, and finding RR Lyrae associated with this galaxy is important to constrain its distance. In this work, we present a search of RR Lyrae in the vicinity of Virgo III based on the time-series $r$-band images taken from the Lulin One-meter Telescope (LOT). We have identified three RR Lyrae from our LOT data, including two fundamental mode (ab-type) and a first-overtone (c-type) RR Lyrae. Assuming these three RR Lyrae are members of Virgo III, we derived the distance to this UFD as $154\pm25$~kpc, fully consistent with the independent measurements given in the literature. We have also revisited the relation between absolute $V$-band magnitude ($M_V$) and the number of RR Lyrae (of all types, $N_{RRL}$) found in local galaxies, demonstrating that the $M_V$-$N_{RRL}$ relation is better described with the specific RR Lyrae frequency.

\end{abstract}


\section{Introduction}\label{sec1}

Finding RR Lyrae in dwarf galaxies, especially the ultra-faint dwarfs \citep[UFD; for a recent review, see][]{simon2019}, is particularly interesting \citep{sesar2014,baker2015}. This is because RR Lyrae are well-known standard candles, therefore distances measured from RR Lyrae can be used to constrain the properties of their host UFD. Recently, \citet{homma2023} reported the discovery of Virgo III as a candidate UFD. Based on the empirical relation derived in \citet{mv2019}, and using the integrated $V$-band absolute magnitude ($M_V$) given in \citet[][with $M_V=-2.69^{+0.45}_{-0.56}$~mag]{homma2023}, the ``expected'' number of RR Lyrae ($N_{RRL}$) in Virgo III is $1\pm1$. Boötes II and Willman 1 have $M_V=-2.9$~mag and $-2.5$~mag, respectively, bracketing Virgo III. Yet Boötes II has one RR Lyrae and Willman 1 has none \citep{tau2024}. Therefore, Virgo III could have (at least) one RR Lyrae or none, and it is useful to search and identify these variables in Virgo III.

In this work, we present our search for potential RR Lyrae in the vicinity of Virgo III using the Lulin One-meter Telescope (LOT), located at the central Taiwan. We first describe our time-series observations carried out at LOT, as well as the image reduction and photometric calibration, in Section \ref{sec2}. We then create a set of simulated light-curves to evaluate the feasibility of detecting RR Lyrae based on the characteristics of our LOT observations, and search for potential RR Lyrae using the calibrated light curves in Section \ref{sec3} and \ref{sec4}, respectively. In Section \ref{sec5} we present our detected RR Lyrae, and revisit the $M_V$-$N_{RRL}$ relation in Section \ref{sec6}. We concluded our work in Section \ref{sec7}.

\section{LOT Observations, Reduction, and Calibration} \label{sec2}

\begin{deluxetable}{ccrcc}
  \tabletypesize{\scriptsize}
  \tablecaption{Log of LOT observations.\label{tab1}}
  \tablewidth{0pt}
  \tablehead{
    \colhead{Date} &
    \colhead{MJD} &
    \colhead{$\Delta t$\tablenotemark{a}} &
    \colhead{FWHM\tablenotemark{b}} &
    \colhead{Depth\tablenotemark{c}} 
  }
  \startdata
2023-12-24 & 60302.796701 &  0.000 & 2.96 & 21.9 \\
2023-12-25 & 60303.839954 &  1.043 & 2.97 & 22.0 \\
2024-01-10 & 60319.845984 & 17.049 & 1.98 & 23.2 \\
           & 60319.860313 & 17.064 & 2.17 & 23.1 \\
           & 60319.874363 & 17.078 & 2.43 & 23.0 \\
2024-01-11 & 60320.735174 & 17.939 & 2.17 & 22.8 \\
           & 60320.749248 & 17.953 & 2.32 & 22.8 \\
           & 60320.763287 & 17.967 & 2.17 & 22.9 \\
2024-01-13 & 60322.871262 & 20.075 & 1.49 & 23.4 \\
           & 60322.885324 & 20.089 & 1.51 & 23.3 \\
2024-01-14 & 60323.847639 & 21.051 & 1.76 & 23.3 \\
           & 60323.861701 & 21.065 & 1.82 & 23.3 \\
           & 60323.875741 & 21.079 & 1.89 & 23.3 \\
           & 60323.890775 & 21.094 & 1.91 & 23.2 \\
2024-01-15 & 60324.849537 & 22.053 & 2.36 & 23.0 \\
           & 60324.863588 & 22.067 & 2.25 & 23.1 \\
           & 60324.877639 & 22.081 & 2.05 & 23.2 \\
           & 60324.891678 & 22.095 & 2.00 & 23.1 \\
2024-03-08 & 60377.669803 & 74.873 & 1.85 & 23.1 \\
           & 60377.869294 & 75.073 & 2.80 & 21.9 \\
           & 60377.883345 & 75.087 & 2.77 & 21.5 \\
\enddata
\tablenotetext{a}{Time-difference, in days, between a given image and the first image.}
\tablenotetext{b}{Averaged full-width at half-maximum (FWHM) of the point sources in the image in unit of arc-second.}
\tablenotetext{c}{The $5\sigma$ limiting magnitude was adopted to represent the depth of each image.} 
\end{deluxetable}

LOT is a $F/8$ Cassegrain reflector, and it was equipped with the Andor iKon-L 936 CCD imager during our queued observations. As a result, the LOT images have a pixel scale of $0.345\arcsec$pixel$^{-1}$ and a field-of-view (FOV) of $11.8\arcmin \times 11.8\arcmin$. Note that the half-light radius for Virgo III is $r_h=1\arcmin$ \citep{homma2023}, therefore the FOV of LOT can cover the entire galaxy. Given the expected faintness of the RR Lyrae ($r\sim21.5$~mag), we only observed Virgo III using the $r$-band filter commercially available from Astrodon, with exposure time of 1200~s (except the first two nights in 2023, when the exposure time was set to 900~s). Log of our time-series observations is given in Table \ref{tab1}.

All of the collected images were bias-subtracted and dark-subtracted using the master-bias and master-dark frames acquired from the same night, followed by flatfielding using either dome flat or twilight flat images. Astrometric calibration on the reduced images were done using the {\tt SCAMP} \citep{scamp2006} software. For photometric calibration, we selected $\sim 20$ reference stars from the Pan-STARRS1 (PS1) photometric catalog \citep{chambers2016,flewelling2020}. Criteria for selecting the PS1 reference stars (whenever applicable) were same as in \citet{ngeow2022} and \citet{ngeow2024}, and hence will not be repeated here. The $r$-band magnitudes and colors of the PS1 reference stars, $r^{PS1}$ and $(g-r)^{PS1}$ respectively, were then used to iteratively fit the regression in the following form:

\begin{eqnarray}
  r^{PS1}-r^{\mathrm{instr}} & = & ZP + C (g-r)^{PS1}.
\end{eqnarray}

\noindent The instrumental magnitudes of the reference stars, $r^{\mathrm{instr}}$ on each images, were based on the point-spread-function (PSF) photometry measured from using the {\tt Source-Extractor} \citep{bertin1996} and {\tt PSFEx} \citep{bertin2011} package. After solving equation (1), the detected sources in each images were calibrated to the PS1 AB magnitude system. We then fitted a low-order polynomial to the calibrated $r$ vs. $\sigma_r$ plot, and estimated the $5\sigma$ limiting magnitude from the fitted polynomial (these fitted polynomials would be used in the light-curve simulations as described in the next Section). They are listed in the last column of Table \ref{tab1}, and most of the images can reach to a nominal depth of $r\sim23$~mag. We have also estimated the expected photometric error at $r=21.5$~mag, which has a median of $\sim 0.05$~mag.

\section{Light Curves Simulations} \label{sec3}

\begin{figure*}
  \epsscale{1.1}
  \plottwo{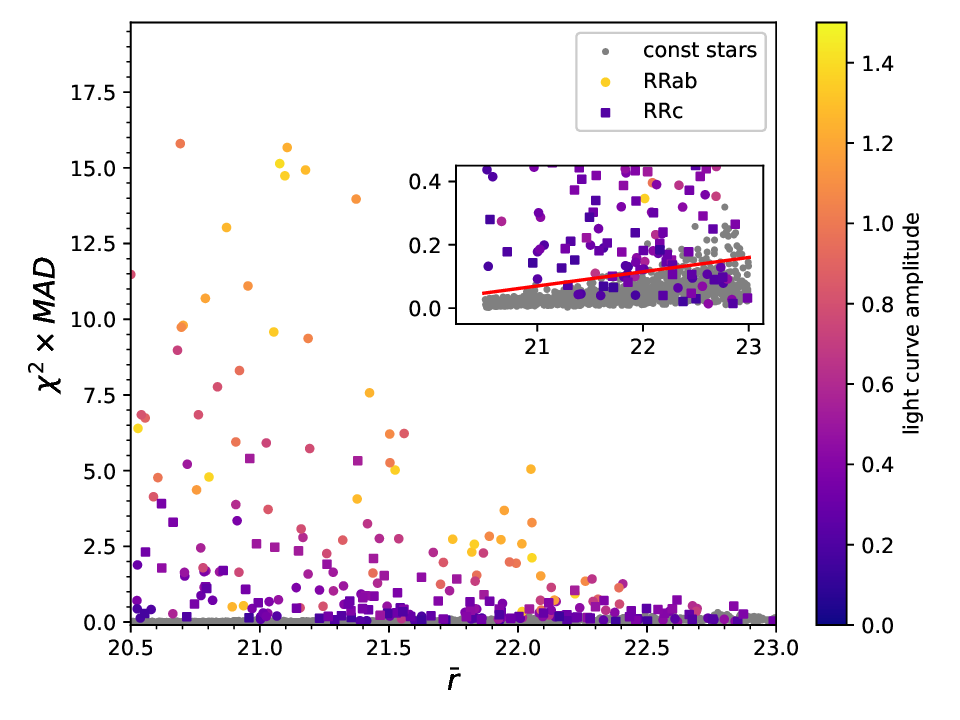}{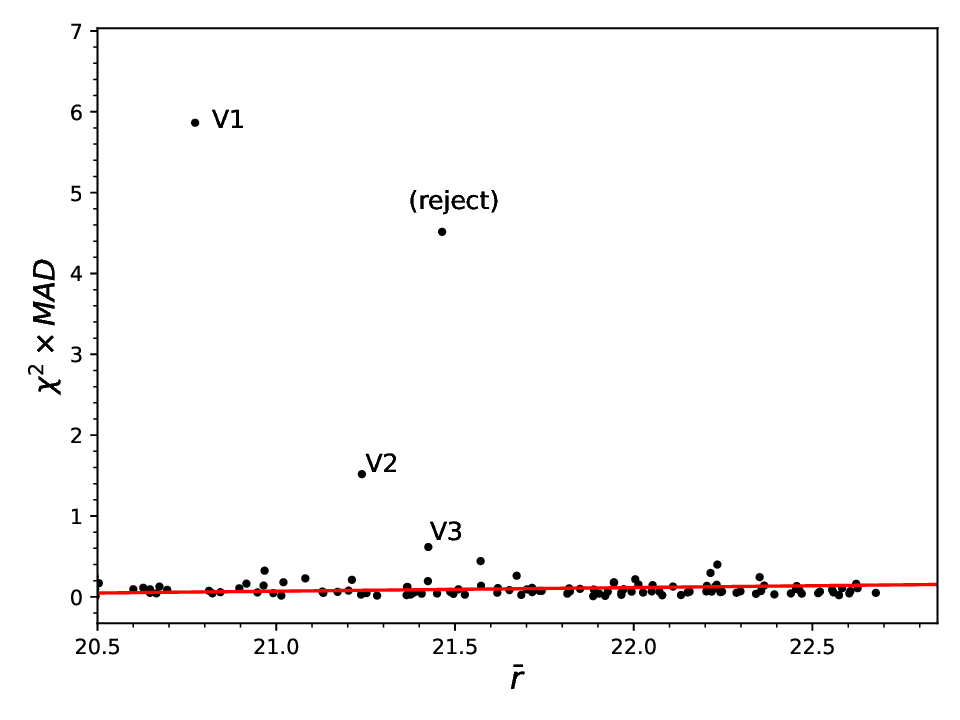}
  \caption{Products of $\chi^2$ and $MAD$ as a function of mean $r$-band magnitudes. The left panel is for simulated light-curves, including constant stars (gray points), ab-type RR Lyrae (circles) and c-type RR Lyrae (squares). The color bar represents the input amplitudes for the simulated light-curves. The inset figure is the zoomed-in version to highlight constant stars. The red line, given as $\chi^2\times MAD=0.045\ r -0.875$, is adopted to separate RR Lyrae and constant stars. The right panel is for those stars, classified in SDSS DR16, with LOT light-curves. The red line is same as the line shown in the inset figure on the left panel. We identified three RR Lyrae labeled as V1, V2, and V3. Light-curve for the star with $\chi^2\times MAD\sim 4.5$ was affected by outliers, and hence it was rejected as a RR Lyrae candidate.}
  \label{fig_var}
\end{figure*}

Given the small number of epochs ($\sim20$) collected from LOT, and our targeted RR Lyrae are faint and close to the detection limit, we ran light-curve simulations to evaluate the feasibility of detecting RR Lyrae using our LOT data.

\subsection{Constant Stars}

We first simulated light-curve for 1000 constant stars based on the epochs as listed in the third column of Table \ref{tab1} ($\Delta t$). The $r$-band magnitudes were uniformly drawn from interval between 20.5~mag and 23.0~mag. Since we expect the RR Lyrae will have $r\sim21.5$~mag, we set the upper limit to be 1~mag brighter than this. The lower limit of 23.0~mag was set by the nominal depth of our LOT images, as we won't be able to detect any stars fainter than this limit. For each drawn magnitudes at a given epoch, we added a Gaussian uncertainty based on the fitted low-order polynomial mentioned in Section \ref{sec2}, to the simulated magnitudes. We discarded the simulated magnitudes if such magnitudes were fainter than the depth at a given epoch as listed in the last column of Table \ref{tab1}. Therefore, some simulated light-curves would have less data-points than others. 

We calculated the following two quantities on the simulated light-curves with more than 10 data-points:\footnote{Only less than 1\% of the light-curves did not fulfill this condition.} $\chi^2=\frac{1}{(N-1)}\sum_{i=1}^N (r_i - \bar{r})^2/\sigma_r^2$ and $MAD =\mathrm{median}(|r_i - \tilde{r}|)$, where $\bar{r}$ and $\tilde{r}$ are the weighted means and medians for $r_i$, respectively. The product of these two quantities appeared to be a good metric to identify  (large-amplitude) RR Lyrae against constant stars \citep{ngeow2020}. The gray points in the left panel of Figure \ref{fig_var} show the distribution of $\chi^2\times MAD$ as a function of magnitudes for the simulated constant stars. As expected, $\chi^2\times MAD$ become larger at fainter magnitudes due to the increasing of photometric uncertainties. Nevertheless, values of $\chi^2\times MAD$ did not exceed 0.3 for the simulated constant stars.

\subsection{RR Lyrae}

Light-curve simulations for RR Lyrae are similar to the constant stars, except an additional step of adopting the $r$-band template light-curves ($T_r$) available from \citet{sesar2010}. In brief, light-curve for a RR Lyrae was constructed using the following expression:

\begin{eqnarray}
  r(\Delta t_i) & = & \bar{r} + AMP_r \times T_r(\Delta t_i - t_0, P),
\end{eqnarray}

\noindent where $AMP_r$, $t_0$, and $P$ are the $r$-band light-curve amplitude, epoch at the maximum light, and pulsation period, respectively. For each RR Lyrae, we generated a uniformly distributed random number for $\bar{r}$, $P$, $t_0$, and $AMP_r$. The range for $\bar{r}$ is same as in the cases of constant stars (20.5~mag to 23.0~mag), while for $t_0$ we set its range to be $-P$ and $0$ (days). Finally, a Gaussian uncertainty, based on the polynomial fits for each of the epochal photometry, was added to $r(\Delta t_i)$ in equation (2). Same as in the cases of constant stars, we discarded $r(\Delta t_i)$ if it is fainter than the depth on each epoch.

\begin{figure}
  \epsscale{1.1}
  \plotone{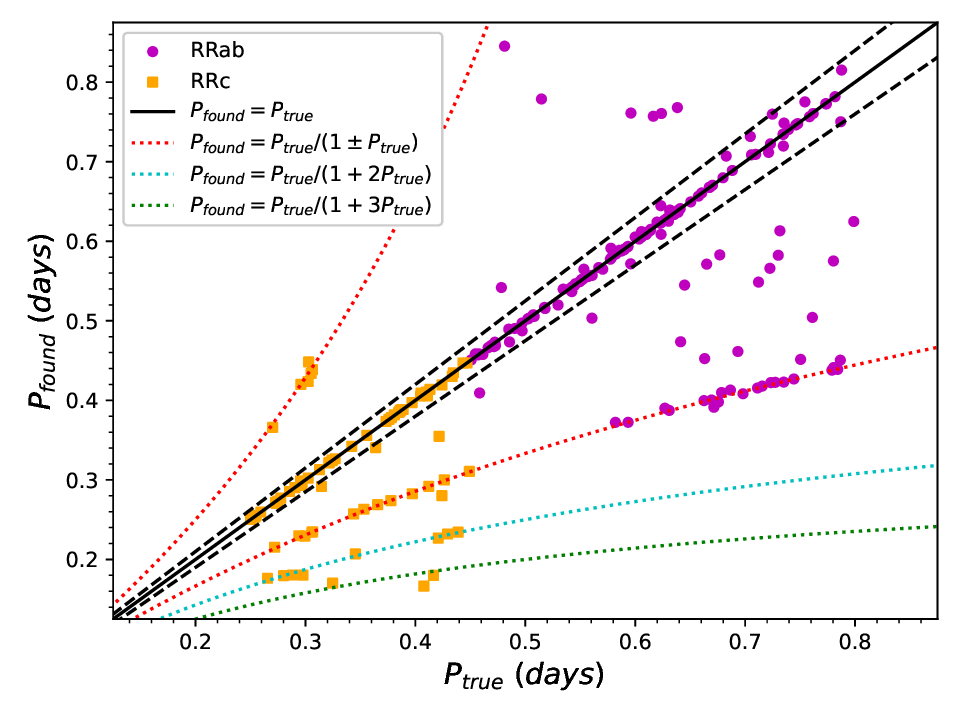}
  \caption{Comparison of the input (true) periods and the periods found, using the {\tt gatspy} package, for our simulated RR Lyrae light-curves. The black dashed lines are for the cases of $P_{\mathrm{found}} = (\pm 5\%) P_{\mathrm{true}}$. The various colors dotted curves represent the aliasing periods. }
  \label{fig_persim}
\end{figure}

RR Lyrae can pulsates either in fundamental or first-overtone mode, known as ab-type and c-type, respectively.\footnote{The third type of RR Lyrae is called d-type, pulsating simultaneously in fundamental and first-overtone mode. However, they are very rare and hence not considered in this work.} Both types of RR Lyrae follow a different distribution of $P$ and $AMP_r$. The adopted ranges for $P$ (in days) are $P_{ab} =[0.45,\ 0.80]$ and $P_{c} =[0.25,\ 0.45]$, while for the $r$-band amplitudes, the adopted ranges are $AMP_{ab}=[0.15,\ 1.40]$ and $AMP_{c}=[0.13,\ 0.55]$ \citep{ngeow2022a}. For template light-curves, 20 and 2 $r$-band templates were available from \citet{sesar2010} for the ab- and c-type, respectively. They were randomly selected when constructing the simulated light-curves via equation (2).

We simulated light-curves for 200 ab-type and 100 c-type RR Lyrae. Values of $\chi^2\times MAD$ for those light-curves with more than 10 data-points were over-plotted in the left panel of Figure \ref{fig_var} alongside with the constant stars. As can be seen from this plot, the (large amplitude) RR Lyrae and constant stars can be well separated using $\chi^2\times MAD$, except for some low-amplitude RR Lyrae toward the faint end (due to larger photometric errors that are comparable to the light-curve amplitudes). The red line shown in the inset figure is a good compromise to separate the RR Lyrae and constant stars, and there were 90\% and 84\% of the simulated ab-type and c-type RR Lyrae, respectively, located above the red line. Hence, the overall completeness of our LOT observations is $\sim 87\%$.

Our simulated RR Lyrae light-curves can also be used to evaluate the period recovery rate. We employed a combination of Lomb-Scargle based and template light-curve period search methods, both implemented in the {\tt gatspy} \citep{vdp2015} package, to search for the periods on our simulated light-curves. We emphasized that these period-search approaches are same as in the search of RR Lyrae in Virgo III using the real LOT data. Figure \ref{fig_persim} presents the result on the period-search, which shows the recovery of 66.1\% of the input periods (indicated as dashed lines in Figure \ref{fig_persim}). Other periods tend to lie along the tracks for different aliasing periods.

\section{Searching for RR Lyrae} \label{sec4}

\begin{deluxetable}{lccc}
  \tabletypesize{\scriptsize}
  \tablecaption{LOT light-curves for the detected RR Lyrae.\label{tab2}}
  \tablewidth{0pt}
  \tablehead{
    \colhead{MJD} &
    \colhead{V1} &
    \colhead{V2} &
    \colhead{V3} 
  }
  \startdata
60324.863588 & $20.991\pm0.031$ & $21.355\pm0.042$ & $21.312\pm0.040$ \\
60319.845984 & $20.893\pm0.027$ & $21.342\pm0.039$ & $21.447\pm0.043$ \\
60377.883345 & $20.525\pm0.086$ & $\cdots$         & $\cdots$ \\
60320.735174 & $21.029\pm0.041$ & $21.347\pm0.054$ & $21.562\pm0.066$ \\
60377.869294 & $20.593\pm0.060$ & $21.006\pm0.087$ & $\cdots$ \\
60324.849537 & $20.985\pm0.032$ & $21.460\pm0.048$ & $21.305\pm0.042$ \\
60324.891678 & $20.990\pm0.029$ & $21.176\pm0.034$ & $21.271\pm0.037$ \\
60303.839954 & $20.666\pm0.062$ & $21.368\pm0.118$ & $\cdots$ \\
60322.871262 & $20.961\pm0.022$ & $21.062\pm0.024$ & $21.480\pm0.034$ \\
60302.796701 & $20.823\pm0.079$ & $\cdots$         & $21.341\pm0.127$ \\
60320.763287 & $20.820\pm0.030$ & $21.433\pm0.052$ & $21.532\pm0.057$ \\
60377.669803 & $20.969\pm0.030$ & $21.431\pm0.045$ & $21.605\pm0.053$ \\
60324.877639 & $21.035\pm0.029$ & $21.320\pm0.038$ & $21.227\pm0.035$ \\
60319.874363 & $21.048\pm0.038$ & $21.478\pm0.054$ & $21.641\pm0.063$ \\
60323.847639 & $20.509\pm0.017$ & $21.295\pm0.032$ & $21.522\pm0.041$ \\
60323.861701 & $20.499\pm0.018$ & $21.326\pm0.034$ & $21.424\pm0.038$ \\
60320.749248 & $20.837\pm0.034$ & $21.455\pm0.059$ & $21.540\pm0.064$ \\
60323.875741 & $20.604\pm0.018$ & $21.379\pm0.034$ & $21.503\pm0.039$ \\
60323.890775 & $20.623\pm0.020$ & $21.413\pm0.039$ & $21.365\pm0.038$ \\
60319.860313 & $20.984\pm0.032$ & $21.372\pm0.043$ & $21.558\pm0.052$ \\
60322.885324 & $20.931\pm0.022$ & $20.910\pm0.022$ & $21.470\pm0.036$ \\
  \enddata
\end{deluxetable}

Since majority of our LOT images can reach to a depth similar to, or slightly deeper than, the Sloan Digital Sky Survey (SDSS) Data Release 16 (DR16) catalog \citep[][at $r\sim22.7$~mag]{sdss2020}\footnote{\url{https://live-sdss4org-dr16.pantheonsite.io/imaging/other_info/}}, and sources in SDSS DR16 have been classified into either stars or galaxies, we adopted SDSS DR16 catalog as our master catalog. There are 226 stellar sources in SDSS D16 located within the footprint of our LOT images, these stellar sources were used to construct a master stars list. We then cross-matched the detected and calibrated sources in each LOT images with this master stars list to create light curves for all of the 226 stellar sources. 

We searched for potential RR Lyrae among 194 light curves that have more than 10 data-points. Values of $\chi^2\times MAD$ for them as a function of mean magnitudes are shown in the right panel of Figure \ref{fig_var}. We visually inspected the light curves and ran a preliminary Lomb-Scargle periodogram analysis for stars above the red line drawn in the right panel of Figure \ref{fig_var}. We identified three RR Lyrae because their periods, amplitudes and folded light-curves resembling a typical RR Lyrae. Two of them, V1 (SDSS objID = 1237654879650775304) and V2 (SDSS objID = 1237654879650841005), are ab-type RR Lyrae, and V3 (SDSS objID = 1237654879650775856) is a c-type RR Lyrae.  LOT $r$-band light-curves for them are presented in Table \ref{tab2}.

\begin{figure}
  \epsscale{1.1}
  \plotone{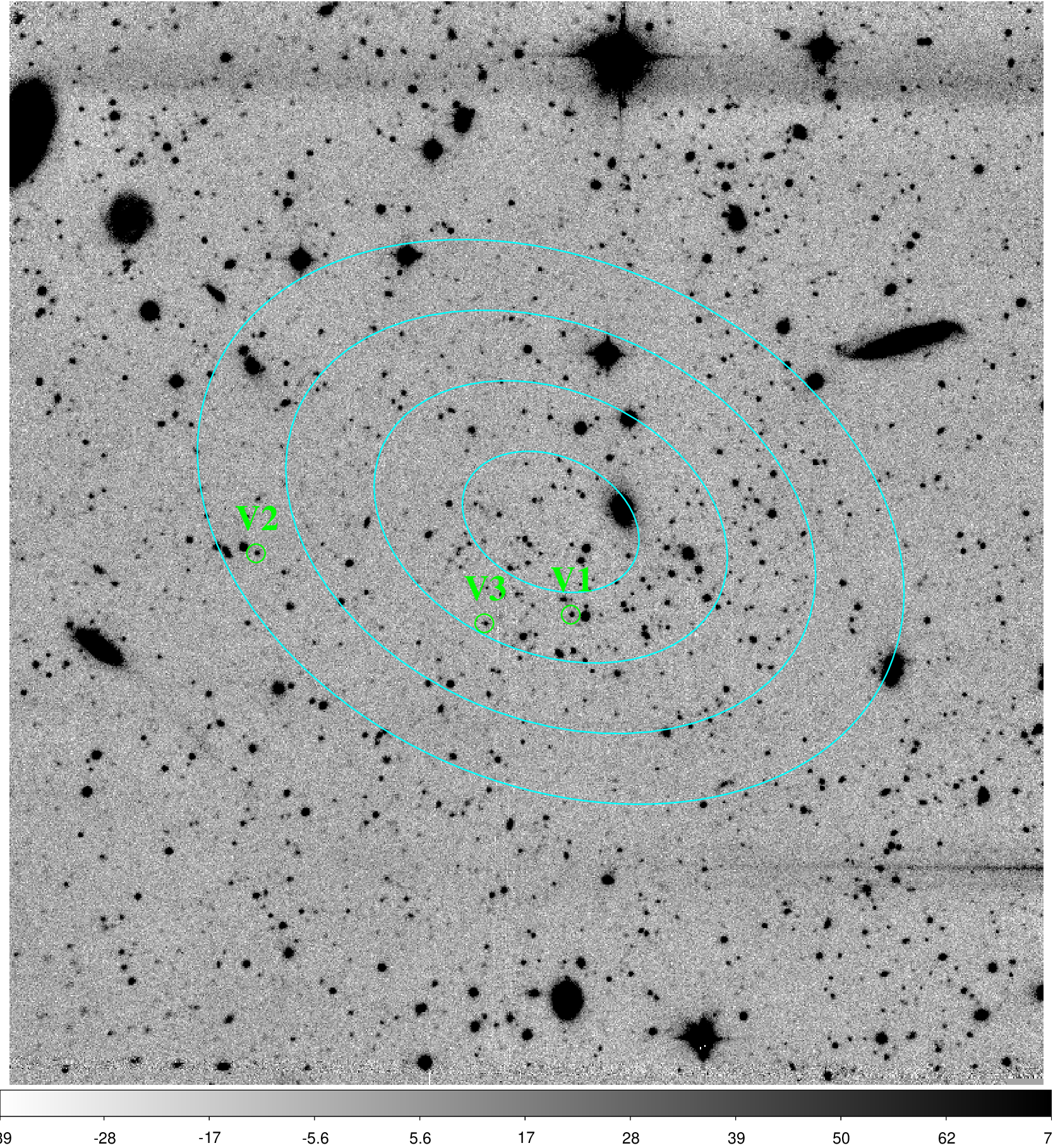}
  \caption{A deep-stack image created by median-combined all of the LOT images using {\tt SWARP} \citep{bertin2002}. The cyan ellipses, centered at $(\alpha,\ \delta)_{J2000}^{\mathrm{Virgo\ III}}$, have semi-major axes of $\{1,\ 2,\ 3,\ 4\}\times r_h$, a projected ellipticity $\epsilon=1-a/r_h$ (where $a$ is the projected semi-minor axis, and $r_h$ is the half-light radius in arc-minute), and with a position angle of $\theta$. Values for $(\alpha,\ \delta)_{J2000}^{\mathrm{Virgo\ III}}$, $r_h$, $\epsilon$, and $\theta$ are all adopted from \citet{homma2023}. Locations of the three detected RR Lyrae were also shown in this deep-stack image.}   
  \label{fig_coadd}
\end{figure}

\begin{figure*}
  \epsscale{1.1}
  \plottwo{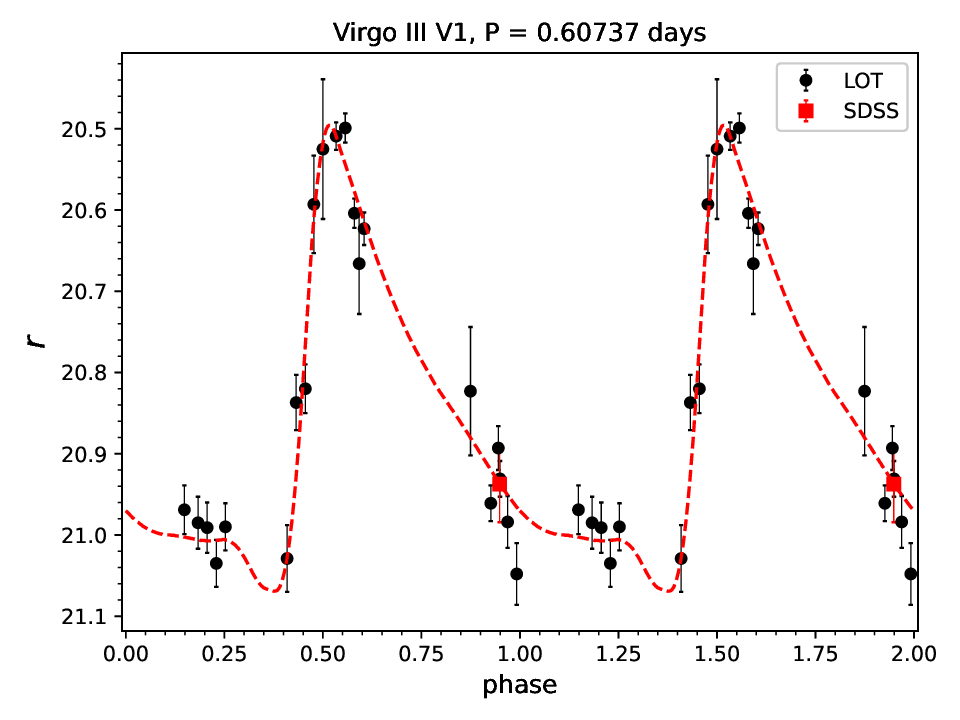}{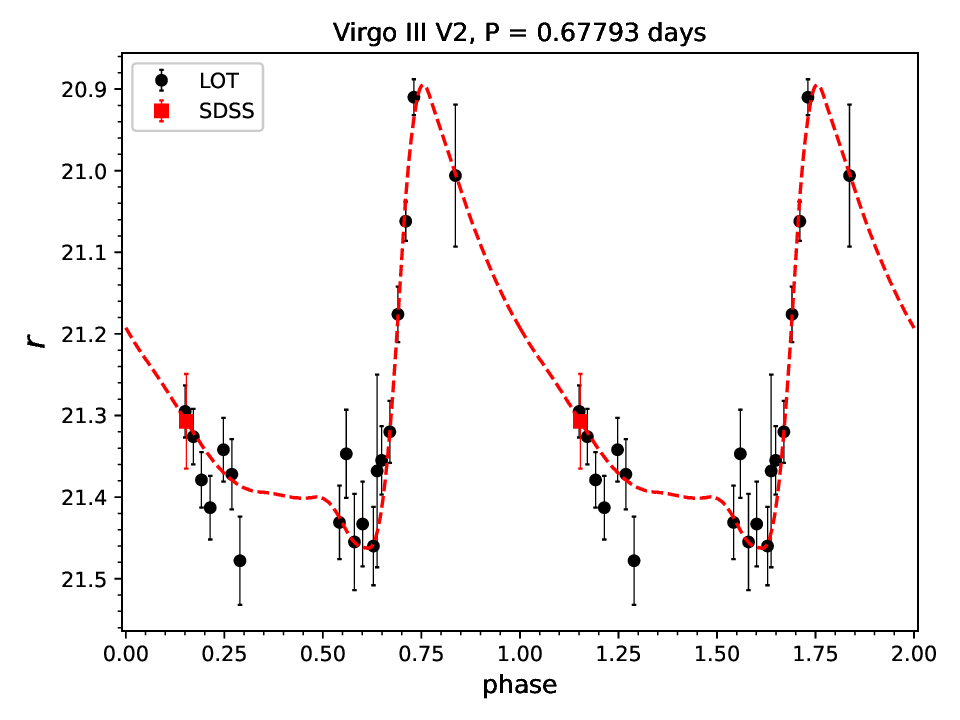}
  \caption{Folded $r$-band light-curves for the two ab-type RR Lyrae V1 and V2. The dashed curves are the best-fit template light-curves returned from the {\tt gatspy's periodic.RRLyraeTemplateModeler} module. The template light-curves were adopted from \citet{sesar2010}.}
  \label{fig_ab}
\end{figure*}

\begin{figure}
  \epsscale{1.1}
  \plotone{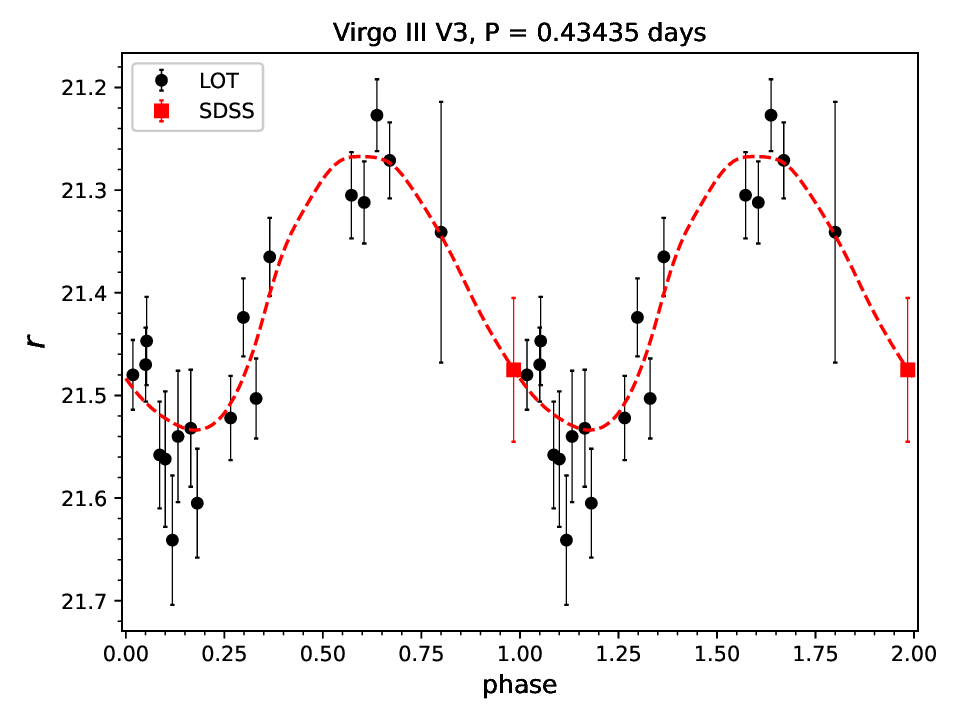}
  \caption{Same as Figure \ref{fig_ab}, but for the c-type RR Lyrae V3.}
  \label{fig_c}
\end{figure}

Figure \ref{fig_coadd} shows the location of the three detected RR Lyrae with respect to Virgo III. The foreground reddening returned from the {\tt Bayerstar2019} 3D reddening map \citep{green2019}\footnote{\url{http://argonaut.skymaps.info/}}, using the {\tt dustmaps} \citep{green2018}\footnote{\url{https://dustmaps.readthedocs.io/en/latest/}} package, toward Virgo III is $E=0.002\pm0.002$~mag. This translates to an $r$-band extinction of $A_r=2.617E=0.005\pm0.005$~mag.

\section{Properties of the Detected RR Lyrae} \label{sec5}

\begin{figure}
  \epsscale{1.1}
  \plotone{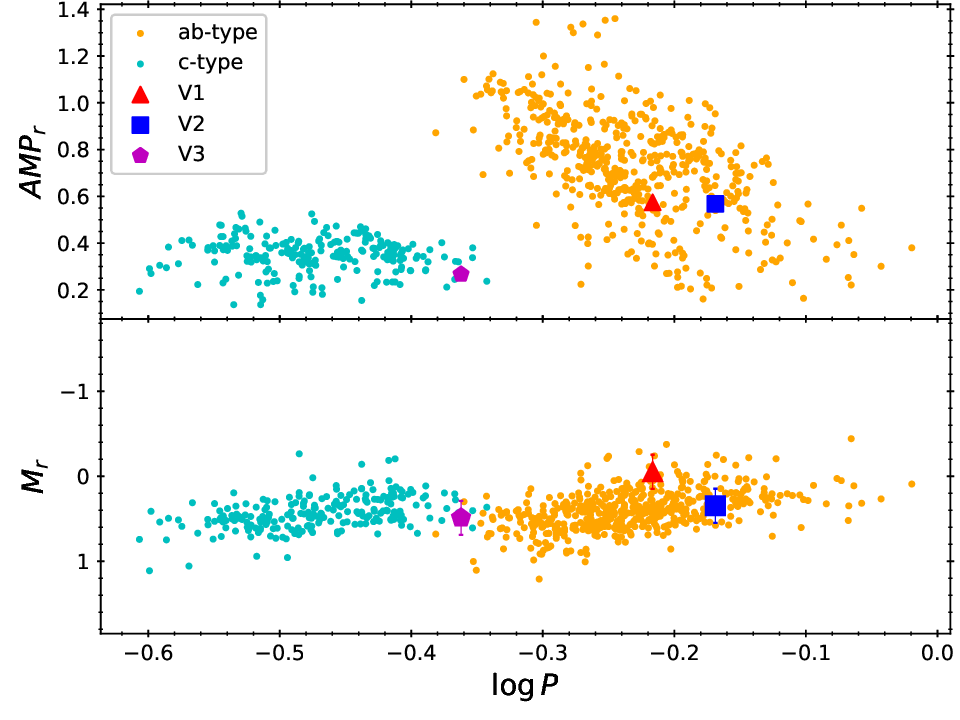}
  \caption{Comparison of $r$-band amplitudes (upper panel) and absolute magnitudes (lower panel) of the three detected RR Lyrae in Virgo III to the RR Lyrae in the globular clusters \citep[orange and cyan symbols, adopted from][]{ngeow2022a}.}
  \label{fig_compare}
\end{figure}

To improve the period determination on the detected RR Lyrae, we added single-epoch SDSS DR16 $r$-band PSF photometry (see Table \ref{tab3}) to the LOT light curves, after converting the SDSS photometry to the PS1 photometric system using the transformation provided in \citet{tonry2012}. We further employed the template light-curve based period search algorithm, available in the {\tt gatspy} package \citep{vdp2015}, for periods refinements. The improved and final adopted periods for the three RR Lyrae are listed in Table \ref{tab3}. Figure \ref{fig_ab} and \ref{fig_c} present the folded light-curves for the ab-type and c-type RR Lyrae, respectively. The dashed curves in these two figures are the best-fitted template light-curves found by {\tt gatspy}, and subsequently used to determine the $r$-band amplitudes and intensity mean magnitudes $\langle r \rangle$. The determined values are listed in Table \ref{tab3}. Figure \ref{fig_compare} compares the $r$-band amplitudes and the extinction-corrected absolute $r$-band magnitude \citep[$M_r$, by adopting the distance modulus, $\mu$, given in][]{homma2023} for the three detected RR Lyrae with the counterparts in the globular clusters \citep{ngeow2022a}. All of the three RR Lyrae are located within the distributions of known RR Lyrae, strongly supporting their identification and membership to Virgo III.

\begin{deluxetable}{lccc}
  \tabletypesize{\scriptsize}
  \tablecaption{Basic properties of the detected RR Lyrae.\label{tab3}}
  \tablewidth{0pt}
  \tablehead{
    \colhead{} &
    \colhead{V1} &
    \colhead{V2} &
    \colhead{V3} 
  }
  \startdata
  & \multicolumn{3}{c}{From SDSS DR16} \\
  $\alpha_{J2000}\ (^\circ)$ & 186.34445301     & 186.40158543 & 186.36018615 \\
  $\delta_{J2000}\ (^\circ)$ & +04.42413144     & +04.43515192 & +04.42253358 \\
  $MJD_{SDSS}$ (days)      & 51987.3194       & 51987.3198   & 51987.3194 \\
  $r_{SDSS}$ (mag)         & $20.938\pm0.047$ & $21.310\pm0.058$ & $21.476\pm0.070$ \\
  & \multicolumn{3}{c}{From LOT light-curve} \\
  Type                    & ab               & ab           & c \\
  $P$ (days)              & 0.60737 & 0.67793  &  0.43435 \\
  $AMP_r$  (mag)          &   0.574   &     0.568  &  0.267 \\
  $\langle r \rangle$ (mag)\tablenotemark{a} & $20.852\pm0.010$ & $21.247\pm0.014$  & $21.394\pm0.016$ \\
  $\mu_r$ (mag)             & $20.542\pm0.305$ & $20.994\pm0.253$  & $21.142\pm0.247$ \\
  \enddata
  \tablenotetext{a}{Errors on $\langle r \rangle$, $\sigma_{\langle r \rangle}$, were estimated using the empirical relation between $\sigma_{\langle r \rangle}$ and number of data-points on light-curves \citep{ngeow2022a}.}
\end{deluxetable}

Photometric metallicity, \feh, for the two ab-type RR Lyrae can be estimated using the empirical relation derived in \citet{saranjedini2006}: \feh~$=-3.43-7.82\log P_{ab}$, which is in the \citet[][ZW84]{zw1984} scale. This empirical relation carries an RMS of $0.45$~dex. Based on RR Lyrae in the globular clusters, \citet{sg2023} demonstrated that this empirical relation can reach to an accuracy of $\pm0.28$~dex. After calculating \feh~for the two ab-type RR Lyrae, we converted the \feh~to the \citet[][D16]{dias2016} scale\footnote{The conversion can be derived using the various \feh~scales summarized in \citet{ngeow2024}, that is: \feh$_{D16} = 1.116$\feh$_{ZW84}+0.212$.} and obtained $[\mathrm{Fe}/H]_{V1}=-1.73$~dex and $[\mathrm{Fe}/H]_{V2}=-2.14$~dex. We adopted the average, \feh$~=-1.93\pm0.37$~dex \citep[the error is based on the small number statistics, see][p. 202]{dean1951,keeping1962}, as the typical \feh~for the three RR Lyrae and Virgo III. Our value is consistent with the work of \citet{homma2023}, who adopted an isochrone filter at metallicity of $-2.2$~dex to fit the Virgo III color-magnitude diagram (CMD). We did not use the $r$-band relation between \feh, $P$, and Fourier parameter $\phi_{31}$ \citep{ngeow2022} to estimate their photometric metallicity because the low-order Fourier expansion failed to provide reasonable fit to the observed light-curves.

\begin{figure}
  \epsscale{1.1}
  \plotone{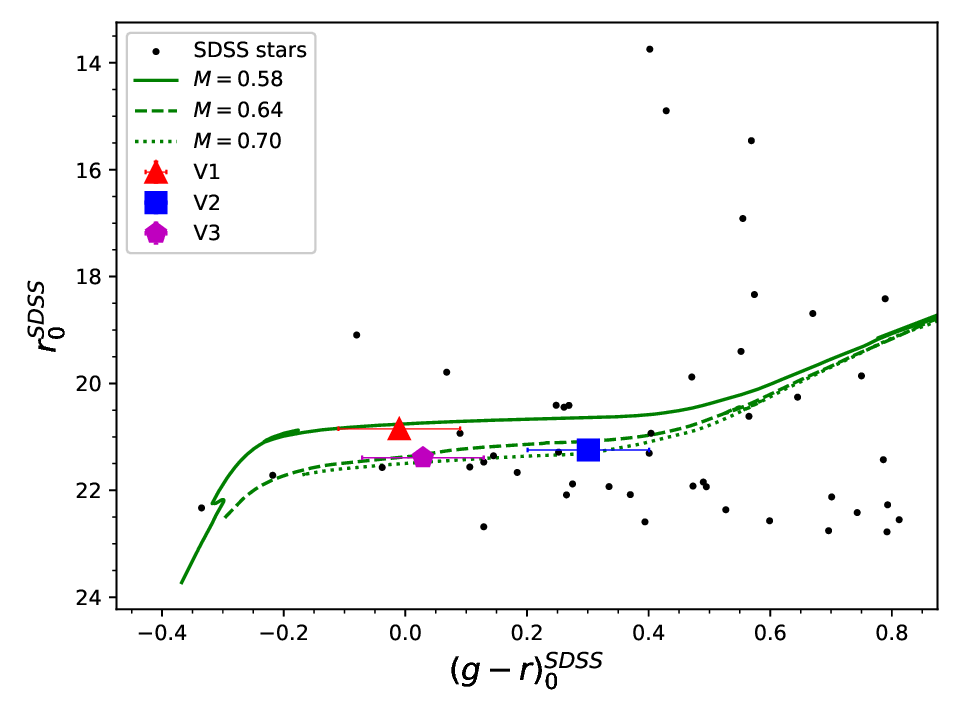}
  \caption{The color-magnitude diagram for stars located within the $4\times r_h$ ellipse of Virgo III, including the three detected RR Lyrae, which have been corrected for extinction. For the three RR Lyrae, we have also applied approximate color corrections to bring the single-epoch SDSS photometry close to the mean values. We assumed such corrections carry an error of $\sim 0.1$~mag. The mean $r$-band magnitudes for them were taken from Table \ref{tab3}. The green curves are three evolutionary tracks for horizontal branch (HB) models with mass ($M$) of $0.58,\ 0.64$, and $0.70$~$M_\odot$, taken from the {\tt BaSTI} library. These HB models are $\alpha$-enhanced \citep[as found in many UFD; for a review, see][]{simon2019}, with \feh$=-1.9$~dex, and vertically shifted to the absolute magnitudes using the distance modulus derived in our work.}   
  \label{fig_cmd}
\end{figure}

Using both of the fundamental mode and first-overtone mode $r$-band period-luminosity-metallicity relations derived in \citet{ngeow2022a}, where the metallicity is in the D16 scale, and the adopted \feh, we calculated the distance moduli for these three RR Lyrae. The results are summarized in the last row of Table \ref{tab3}. By taking a weighted average and assuming all three RR Lyrae are genuine member of Virgo III, the distance modulus for Virgo III was found to be $\mu = 20.937\pm0.355$~mag (statistical error only from small number statistics), or a linear distance of $154\pm25$~kpc. Our distance modulus is fully consistent with the values given in \citet{homma2023}, $20.9\pm0.2$~mag and $20.91\pm0.04$~mag, which were derived using isochrone fitting and the blue horizontal-branch (BHB) stars, respectively.   

In Figure \ref{fig_cmd}, we presented the extinction-corrected CMD for SDSS stars located within the $4\times r_h$ ellipse of Virgo III (see Figure \ref{fig_coadd}). Since the SDSS photometry was based on the single-epoch observations, we applied additional color corrections such that the SDSS photometry are close to the mean $(g-r)$ colors for these RR Lyrae. These approximate corrections were estimated based on the template light curves \citep{sesar2010} or a $(g-r)$ template color curve \citep{ngeow2022}. Overplotted on the CMD are the evolutionary tracks for the horizontal branch (HB) models at three representative masses. These evolutionary tracks were taken from the {\tt BaSTI} \citep[a Bag of Stellar Tracks and Isochrones,][]{basti2021} stellar isochrones library. As can be seen from the CMD, the locations of three detected RR Lyrae are consistent with the HB evolutionary tracks corresponding to our adopted distance modulus to Virgo III, strengthening their identification as RR Lyrae and the membership to the galaxy.

\section{The $M_V$-$N_{RRL}$ Relation} \label{sec6}

\begin{deluxetable*}{lcrl|lcrl}
  \tabletypesize{\scriptsize}
  \tablecaption{Summary of $M_V$ and $N_{RRL}$ for local galaxies.\label{tab4}}
  \tablewidth{0pt}
  \tablehead{
    \colhead{Galaxy} &
    \colhead{$M_V$} &
    \colhead{$N_{RRL}$} &
    \colhead{Reference\tablenotemark{a}} &
    \colhead{Galaxy} &
    \colhead{$M_V$} &
    \colhead{$N_{RRL}$} &
    \colhead{Reference\tablenotemark{a}} 
  }
  \startdata
  Triangulum 	& $-18.8\pm0.1$ & 99 & (15)/(2)                & Andromeda XIII	& $-6.5\pm0.7$ & 9 & (1)/(1) \\
  LMC 	& $-18.1\pm0.1$ & 39082 & (1)/(1)                      & Andromeda XI	& $-6.3\pm0.6$ & 15 & (1)/(1) \\
  SMC 	& $-16.8\pm0.2$ & 6369 & (1)/(1)                       & Boötes I	& $-6.0\pm0.3$ & 27 & (1)/(3)\tablenotemark{b} \\
  NGC 205	& $-16.5\pm0.1$ & 30 & (15)/(2)                & Boötes III	& $-5.8\pm0.5$ & 32 & (1)/(3)\tablenotemark{b} \\
  M32 	& $-16.4\pm0.2$ & 509 & (15)/(2)                       & Hercules 	& $-5.8\pm0.2$ & 12 & (1)/(1) \\
  IC 1613	& $-15.2\pm0.2$ & 90 & (1)/(1)                 & Sagittarius 2	& $-5.7\pm0.1$ & 5 & (1)/(1) \\
  NGC 6822	& $-15.2\pm0.2$ & 26 & (1)/(1)                 & Centaurus 1	& $-5.6\pm0.1$ & 3 & (14)/(2) \\
  NGC 185	& $-14.8\pm0.1$ & 818 & (1)/(16)               & Canes Venatici II	& $-5.2\pm0.3$ & 2 & (1)/(1) \\
  NGC 147	& $-14.6\pm0.1$ & 177 & (1)/(1)                & Ursa Major I	& $-5.1\pm0.4$ & 8 & (1)/(3) \\
  WLM 	& $-14.2\pm0.1$ & 90 & (15)/(20)                       & Leo IV	& $-5.0\pm0.3$ & 3 & (1)/(1) \\
  Fornax 	& $-13.5\pm0.1$ & 2068 & (1)/(6)               & Hydrus I	& $-4.7\pm0.1$ & 5 & (1)/(3) \\
  Sagittarius dSph	& $-13.5\pm0.3$ & 2045 & (1)/(1)       & Eridanus IV	& $-4.7\pm0.2$ & 0 & (9)/(3) \\
  Andromeda VII	& $-13.2\pm0.3$ & 573 & (1)/(1)                & Hydra II	 & $-4.6\pm0.4$ & 1 & (1)/(1) \\
  UGC4879 	& $-12.5\pm0.2$ & 678 & (15)/(2)               & Carina II	& $-4.5\pm0.1$ & 3 & (1)/(3) \\
  Leo A	& $-12.1\pm0.2$ & 10 & (1)/(1)                         & Leo V	& $-4.4\pm0.4$ & 3 & (1)/(1) \\
  Leo I	& $-11.8\pm0.3$ & 164 & (1)/(1)                        & Coma Berenices	& $-4.3\pm0.3$ & 3 & (1)/(3) \\
  Andromeda II	& $-11.6\pm0.2$ & 251 & (1)/(1)                & Pegasus IV	& $-4.3\pm0.2$ & 3 & (10)/(3) \\
  ESO410-G005 	& $-11.5\pm0.3$ & 269 & (1)/(2)                & Aquarius II	& $-4.3\pm0.1$ & 0 & (1)/(3) \\
  Andromeda VI	& $-11.5\pm0.2$ & 111 & (1)/(1)                & Ursa Major II	& $-4.2\pm0.3$ & 6 & (1)/(3) \\
  Cetus 	& $-11.3\pm0.2$ & 630 & (1)/(1)                & Pisces II	& $-4.2\pm0.4$ & 1 & (17)/(2) \\
  Andromeda I	& $-11.2\pm0.2$ & 296 & (1)/(1)                & Pegasus III	& $-4.1\pm0.4$ & 1 & (12)/(2) \\
  ESO294-G010 	& $-11.2\pm0.3$ & 232 & (1)/(1)                & Tucana II	& $-3.9\pm0.2$ & 4 & (4)/(3) \\
  Sculptor 	& $-10.8\pm0.1$ & 536 & (1)/(1)                & Grus II	& $-3.9\pm0.2$ & 2 & (1)/(3)\tablenotemark{c} \\
  Aquarius 	& $-10.6\pm0.1$ & 32 & (1)/(1)                 & Reticulum II	& $-3.9\pm0.4$ & 0 & (17)/(3) \\
  KKR 25	& $-10.5\pm0.2$ & 46 & (15)/(2)                & Grus I	& $-3.5\pm0.6$ & 2 & (1)/(1) \\
  Pisces I	& $-10.1\pm0.1$ & 56 & (15)/(21)               & Cetus III	& $-3.5\pm0.5$ & 1 & (12)/(3) \\
  Phoenix 	& $-9.9\pm0.4$ & 121 & (1)/(1)                 & Horologium I	 & $-3.5\pm0.6$ & 0 & (17)/(3) \\
  Leo II	& $-9.7\pm0.1$ & 140 & (1)/(1)                 & Phoenix II	& $-3.3\pm0.6$ & 1 & (1)/(3) \\
  Tucana 	& $-9.5\pm0.2$ & 358 & (1)/(1)                 & Kim 2/Indus I& $-3.3\pm0.6$ & 0 & (1)/(1) \\
  Andromeda III	& $-9.5\pm0.3$ & 111 & (1)/(1)                 & Reticulum III	& $-3.3\pm0.3$ & 0 & (11)/(3) \\
  Carina 	& $-9.4\pm0.1$ & 92 & (1)/(2)                  & Tucana IV	& $-3.0\pm0.4$ & 0 & (22)/(3)\tablenotemark{d} \\
  Andromeda XXI	& $-9.1\pm0.3$ & 41 & (1)/(1)                  & Boötes II	& $-2.9\pm0.7$ & 1 & (1)/(3) \\
  Leo P	& $-9.1\pm0.2$ & 10 & (1)/(1)                          & Virgo III	& $-2.7\pm0.6$ & 3 & (12)/(25) \\
  Antlia 2	& $-9.0\pm0.2$ & 318 & (23)/(24)               & Willman 1	& $-2.5\pm0.7$ & 0 & (1)/(3) \\
  Ursa Minor	& $-9.0\pm0.1$ & 82 & (1)/(1)                  & Carina III	& $-2.4\pm0.2$ & 0 & (1)/(3) \\
  Andromeda XXV	& $-9.0\pm0.3$ & 56 & (1)/(1)                  & Eridanus III	& $-2.4\pm0.9$ & 0 & (17)/(3) \\
  Andromeda XIX	& $-9.0\pm0.6$ & 31 & (1)/(1)                  & Delve 2	& $-2.1\pm0.5$ & 0 & (8)/(3)\tablenotemark{d} \\
  CanesVenatici I	& $-8.8\pm0.1$ & 23 & (1)/(1)          & Segue 2	& $-1.9\pm0.9$ & 1 & (1)/(3) \\
  Draco 	& $-8.7\pm0.1$ & 336 & (1)/(18)                & Horologium II	& $-1.6\pm1.0$ & 0 & (17)/(3) \\
  Sextans 	& $-8.7\pm0.1$ & 227 & (1)/(1)                 & Tucana III	& $-1.3\pm0.2$ & 6 & (19)/(3) \\
  Andromeda XXVIII	& $-8.7\pm0.4$ & 85 & (1)/(1)          & Segue 1	& $-1.3\pm0.7$ & 1 & (1)/(3) \\
  Crater II	& $-8.2\pm0.1$ & 99 & (1)/(1)                  & Triangulum II	& $-1.2\pm0.4$ & 0 & (7)/(3) \\
  Andromeda XV	& $-8.0\pm0.4$ & 117 & (1)/(1)                 & Tucana V	& $-1.1\pm0.6$ & 0 & (22)/(3) \\
  Andromeda XXVII	& $-7.9\pm0.5$ & 89 & (1)/(1)          & Segue 3	& $-0.9\pm0.7$ & 0 & (17)/(5) \\
  Leo T	& $-7.6\pm0.1$ & 5 & (1)/(2)                           & Virgo I	& $-0.9\pm0.7$ & 0 & (12)/(3) \\
  Andromeda XVI	& $-7.3\pm0.3$ & 8 & (1)/(1)                   & Draco II	& $-0.8\pm1.0$ & 0 & (13)/(3) \\
  Eridanus 2	& $-7.2\pm0.1$ & 67 & (17)/(2)                 & Cetus II	& $0.0\pm0.7$  & 0 & (11)/(3) \\
  \enddata
  \tablenotetext{a}{Left and right entries of the slash are the reference or sources for $M_V$ and $N_{RRL}$, respectively. (1) \citet{mv2019} and reference therein; (2) \citet{monelli2022} and reference therein; (3) \citet{tau2024}; (4) \citet{bechtol2015}; (5) \citet{boettcher2013}; (6) \citet{braga2022}; (7) \citet{carlin2017}; (8) \citet{cerny2021a}; (9) \citet{cerny2021b}; (10) \citet{cerny2023}; (11) \citet{dw2015}; (12) \citet{homma2023}; (13) \citet{longeard2018}; (14) \citet{mau2020}; (15) \citet{mc2012}; (16) \citet{monelli2017}; (17) \citet{munoz2018}; (18) \citet{muraveva2020}; (19) \citet{mp2018}; (20) \citet{sarajedini2023}; (21) \citet{sarajedini2024}; (22) \citet{simon2020}; (23) \citet{torrealba2019}; (24) \citet{vivas2022}; (25) This work.}
  \tablenotetext{b}{RR Lyrae found in these two galaxies could include members from the Sagittarius stream or belong to the (random) halo stars, and the actual number of RR Lyrae could be lower. See \citet{tau2024} for more discussion.}
  \tablenotetext{c}{\citet{tau2024} detected six RR Lyrae, but four of them are brighter and associated with the Chenab/Orphan stream. Hence, there is only two RR Lyrae associated with Grus II.}
  \tablenotetext{d}{As argued in \citet{tau2024}, the 2 and 3 detected RR Lyrae for Delve 2 and Tucana IV, respectively, are probably belong to SMC. Therefore we assigned $N_{RRL}=0$ for both galaxies.}
\end{deluxetable*}

In this Section, we revisited the $M_V$-$N_{RRL}$ (of all types) relation presented in \citet{mv2019}. We began with 63 galaxies listed in the appendix of \citet{mv2019}, and added new local galaxies from \citet{monelli2022} and \citet{tau2024}, supplemented with several additional galaxies (such as Virgo III from this work) not included in these compilations. We have also updated the number of RR Lyrae in a few dwarf galaxies based on the latest publications (in particular, for Fornax and Draco). The updated list of local galaxies, with their $M_V$ and $N_{RRL}$, is provided in Table \ref{tab4}, alongside with the references for $M_V$ and $N_{RRL}$. In total, there are 57867 RR Lyrae found in 94 galaxies, but $\sim78.5\%$ of them come from the Magellanic Clouds.

\begin{figure}
  \epsscale{1.15}
  \plotone{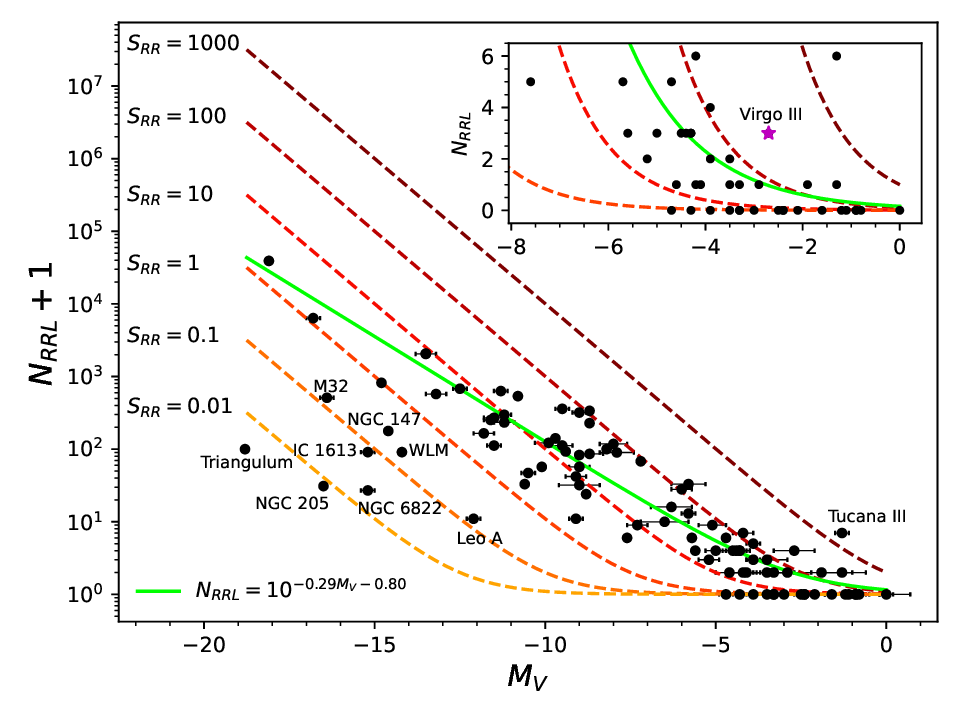}
  \caption{Number of RR Lyrae ($N_{RRL}$) as a function of $M_V$ for the 94 galaxies listed in Table \ref{tab4}. To deal with the ``log of zero'' problem on the $y$-axis (in log-scale), we added $+1$ to the $N_{RRL}$. The green solid curve represents the empirical relation derived in \citet{mv2019}. The dashed curve are for the different selected values of specific RR Lyrae frequency $S_{RR}$. Several ``outliers'' are also marked in this plot.  The inset figure is the zoomed-in version for UFD galaxies with $N_{RRL}<7$, where Virgo III is marked as a magenta star. For clarity, error bars and the curves for $S_{RR}<1$ are omitted in the inset figure. Note that the $y$-axis is in linear scale in the inset figure.}
  \label{fig_nrrl}
\end{figure}

Figure \ref{fig_nrrl} presents the $M_V$-$N_{RRL}$ relation for the galaxies listed in Table \ref{tab4}. The (green) solid line is the empirical relation derived in \citet{mv2019}, which describes the trend of the data well, therefore we did not re-derive the relation. Instead, the relation shows a large scatter. Therefore, we over-plotted the inverted specific RR Lyrae frequency \citep[$S_{RR}$,][]{suntzeff1991,mackey2003} as: 

\begin{eqnarray}
  N_{RRL} & = & S_{RR} 10^{-0.4(M_V+7.5)},
\end{eqnarray}

\noindent which is normalized to $M_V=-7.5$~mag. Figure \ref{fig_nrrl} shows the curves of equation (3) for several representative $S_{RR}$, and most of the galaxies are confined between the curves for $S_{RR}=1$ and $S_{RR}=100$. Furthermore, the local galaxies seems to follow various tracks at given $S_{RR}$ (for examples, LMC, SMC, and NGC 185 are located at the $S_{RR}=1$ track, and some of the $M_V > -10$~mag galaxies are located along the $S_{RR}=100$ track). Clearly, there is no single value of $S_{RR}$ to fit the majority of the galaxies \citep[also, see][]{baker2015}.

Tucana III is the only faint UFD with $S_{RR}>1000$. However, all of the six detected RR Lyrae are extra-tidal stars of Tucana III \citep{vivias2020}. Furthermore, \citet{tau2024} only recovered one of them. It is possible that Tucana III might only have one extra-tidal RR Lyrae, which reduced its $S_{RR}$ to $\sim302$. On the other hand, there are several galaxies below the curve of $S_{RR}=1$, implying the detection of RR Lyrae on these galaxies are not yet completed. This is especially true for Triangulum, as the known RR Lyrae in this galaxy were detected based on the several narrow ``pencil-beam'' fields around Triangulum \citep{tanakul2017}. 

\section{Conclusion} \label{sec7}

In this work, we searched for RR Lyrae in Virgo III using the time-series LOT observations. We have also ran light-curve simulations by taking the characteristics of LOT observations (such as photometric errors and depths on each images) into account, and demonstrated that RR Lyrae can be detected using the LOT data and our searching method (i.e. the $\chi^2 \times MAD$ metric).  We identified two ab-type and one c-type RR Lyrae with periods and amplitudes consistent with the known RR Lyrae in the globular clusters. Given that they are located within the $4\times r_h$ ellipse of Virgo III, and have similar distance modulus as Virgo III, we assume they are true members of Virgo III. Based on the three detected RR Lyrae in Virgo III, together with the latest findings in the literature, we have also revisited the $M_V$-$N_{RRL}$ relation for local galaxies, showing their relations are better described using the specific RR Lyrae frequency.

It is worth to point out that both Virgo III and the three RR Lyrae are fainter than the detection limit of {\it Gaia}, hence there is no proper-motion information to verify the status of their membership. Confirmation of their membership has to wait for the future radial-velocity measurements. We have also estimated the metallicity of Virgo III to be \feh~$=-1.93$~dex based on the two ab-type RR Lyrae. Nevertheless, assuming the three detected RR Lyrae are members of Virgo III, the RR Lyrae-based distance modulus is fully consistent with the independent measurements given in \citet{homma2023}. 
 
\acknowledgments

We thank the useful discussions and comments from an anonymous referee to improve the manuscript. We are thankful for funding from the National Science and Technology Council (NSTC, Taiwan) under the grant 112-2112-M-008-042. We sincerely thank the observing staff at the Lulin Observatory, C.-S. Lin, H.-Y. Hsiao, and W.-J. Hou, for carrying out the queue observations for this work. This publication has made use of data collected at Lulin Observatory, partly supported by NSTC grant 109-2112-M-008-001. This research has made use of the SIMBAD database and the VizieR catalogue access tool, operated at CDS, Strasbourg, France. This research made use of Astropy,\footnote{\url{http://www.astropy.org}} a community-developed core Python package for Astronomy \citep{astropy2013, astropy2018, astropy2022}. 

The Pan-STARRS1 Surveys (PS1) and the PS1 public science archive have been made possible through contributions by the Institute for Astronomy, the University of Hawaii, the Pan-STARRS Project Office, the Max-Planck Society and its participating institutes, the Max Planck Institute for Astronomy, Heidelberg and the Max Planck Institute for Extraterrestrial Physics, Garching, The Johns Hopkins University, Durham University, the University of Edinburgh, the Queen's University Belfast, the Harvard-Smithsonian Center for Astrophysics, the Las Cumbres Observatory Global Telescope Network Incorporated, the National Central University of Taiwan, the Space Telescope Science Institute, the National Aeronautics and Space Administration under Grant No. NNX08AR22G issued through the Planetary Science Division of the NASA Science Mission Directorate, the National Science Foundation Grant No. AST-1238877, the University of Maryland, Eotvos Lorand University (ELTE), the Los Alamos National Laboratory, and the Gordon and Betty Moore Foundation.

Funding for the Sloan Digital Sky Survey IV has been provided by the Alfred P. Sloan Foundation, the U.S. Department of Energy Office of Science, and the Participating Institutions. SDSS acknowledges support and resources from the Center for High-Performance Computing at the University of Utah. The SDSS web site is \url{www.sdss4.org}.

SDSS is managed by the Astrophysical Research Consortium for the Participating Institutions of the SDSS Collaboration including the Brazilian Participation Group, the Carnegie Institution for Science, Carnegie Mellon University, Center for Astrophysics | Harvard \& Smithsonian (CfA), the Chilean Participation Group, the French Participation Group, Instituto de Astrofísica de Canarias, The Johns Hopkins University, Kavli Institute for the Physics and Mathematics of the Universe (IPMU) / University of Tokyo, the Korean Participation Group, Lawrence Berkeley National Laboratory, Leibniz Institut für Astrophysik Potsdam (AIP), Max-Planck-Institut fur Astronomie (MPIA Heidelberg), Max-Planck-Institut für Astrophysik (MPA Garching), Max-Planck-Institut für Extraterrestrische Physik (MPE), National Astronomical Observatories of China, New Mexico State University, New York University, University of Notre Dame, Observatório Nacional / MCTI, The Ohio State University, Pennsylvania State University, Shanghai Astronomical Observatory, United Kingdom Participation Group, Universidad Nacional Autónoma de México, University of Arizona, University of Colorado Boulder, University of Oxford, University of Portsmouth, University of Utah, University of Virginia, University of Washington, University of Wisconsin, Vanderbilt University, and Yale University.

\facility{LO:1m, PS1, Sloan}

\software{{\tt astropy} \citep{astropy2013,astropy2018,astropy2022}, {\tt dustmaps} \citep{green2018}, {\tt gatspy} \citep{vdp2015}, {\tt Matplotlib} \citep{hunter2007},  {\tt NumPy} \citep{harris2020}, {\tt PSFEx} \citep{bertin2011}, {\tt SCAMP} \citep{scamp2006}, {\tt Source-Extractor} \citep{bertin1996}, {\tt SWARP} \citep{bertin2002}}


\end{document}